\documentclass[aip,sd,amsmath,amssymb,reprint]{revtex4-1}
\usepackage{graphicx}
\usepackage{dcolumn}
\usepackage{bm}
\usepackage{graphicx}
\usepackage{amsmath}
\usepackage{mathptm} 
\usepackage{hyperref}
\preprint{APS/123-QED}
\usepackage[normalem]{ulem}
\usepackage{xcolor}

\begin{document}
\preprint{APS/123-QED}

\title{Enhancement of dynamical robustness in a mean-field coupled  network through self-feedback delay}

\author{Amit Sharma}
\email{sharma.a28@gmail.com}
 \affiliation{Department of Physics, Central University of Rajasthan, Ajmer 305 817, India} 
  
\author{Biswambhar Rakshit}%
 \email{biswambhar.rakshit@gmail.com}
\affiliation{Department of Mathematics, Amrita School of Engineering, Coimbatore, Amrita Vishwa Vidyapeetham, $641 112$, India}%

\date{\today}

\begin{abstract}
In this article, we propose a very efficient technique to enhance the dynamical robustness for a network of mean-field coupled oscillators experiencing aging transition. In particular, we present a control mechanism based on delayed negative self-feedback, which can effectively enhance dynamical activities in a mean-field coupled network of active and inactive oscillators. Even for a small value of delay, robustness gets enhanced to a significant level. In our proposed scheme, the enhancing effect is more pronounced for strong coupling. To our surprise even if all the oscillators perturbed to equilibrium mode delayed negative self-feedback able to restore oscillatory activities in the network for strong coupling strength. We demonstrate that our proposed mechanism is independent of coupling topology. For a globally coupled network, we provide numerical and analytical treatment to verify our claim. Also, for global coupling to establish the generality of our scheme, we validate our results for both  Stuart-Landau limit cycle oscillators and chaotic R\"ossler oscillators. To show that our scheme is independent of network topology, we also provide numerical results for the local mean-field coupled complex network.
\end{abstract}
\pacs{Valid PACS appear here}
\keywords{Dynamical Robustness, Aging transition, Self-feedback delay, Coupled Oscillators}
                           
\maketitle

\section{Introduction}

Exploring emergent dynamics of a vast network of coupled oscillators got considerable attention in recent years due to its applicability to understanding various self-organized complex systems \cite{Kurths-book,Strogatz,kuramoto1984}. The emergent behaviors of such complex systems depend on the individual dynamics of local sub-units as well as the network topology. The normal functioning of many natural systems requires stable and robust oscillatory dynamics. Therefore, the macroscopic dynamics of such a large-scale system should exhibit resilience against local perturbations.

In recent times, much attention has been invested in understanding the dynamical robustness, which is defined as the ability of a network of oscillators to regulate its rhythmic  activity when a fraction of the dynamical units is malfunctioning\cite{tanaka2012sr}. Physically this situation can be modeled as a network of coupled oscillators where oscillatory nodes switch to equilibrium mode progressively \cite{daido2004}. If the number of nodes which perturbed to equilibrium mode reaches a critical level, the normal activities of such systems may hamper and face severe disruption. This emergent phenomenon is described as aging transition \cite{Tanaka2014,Tanaka2015,Tanaka2015a,Tanaka2017,daido2004,tanaka2012sr,liu2016,tanaka2010p,daido2007,PhysRevERakshit17}. The aging transition might have catastrophic effects in many natural and real-world systems such as metapopulation dynamics in ecology, neuronal dynamics in brain, cardiac oscillations, and power-grid network \cite{Gilarranz2012,Ranta2008,Lisman2008,Frasaca2017}. Therefore, it is of great practical significance to propose some remedial measures or control mechanisms to enhance the dynamical robustness of the coupled systems against the aging or deterioration of the individual unit. Till now, researchers mainly studied aging transition not only considering different network topologies but also using different coupling functions \cite{tanaka2012sr,Srilena2019_nonlinear_dyna,Ray_2020}. Some recent efforts have been directed to explore the possible mechanisms to enhance the dynamical robustness to avoid catastrophic transition. {Liu et al. \cite{liu2016} proposed a mechanism for robustness enhancement that involves an additional parameter to control the diffusion rate.  Kundu et al. \cite{Srilena1,Srilena2} have shown that the robustness can be enhanced by a positive feedback mechanism as well as asymmetric couplings. Bera\cite{bidesh2019} has employed low pass filtering mechanism for enhancing dynamical robustness. Despite such attempts, enhancement mechanisms are yet to be explored fully and deserve significant attention. 
            
All the previous studies on the aging transition considered  nearest neighbour diffusive coupling, which effectively depicts many real-world complex systems. However, for many biological and physical systems, the mean-field coupling is also relevant.  It has been shown that the diffusion process for a network of genetic oscillators can be modelled as mean-field coupling through a regulation known as quorum-sensing. \cite{PRL_Garcia-Ojalvo,NAS_Garcia-Ojalvo}. A network of mean-field coupled oscillators can model the dynamics of SCN neurons\cite{Gonze2005}, which produces circadian oscillations. In this article, we explore a possible mechanism to augment the dynamical robustness of a mean-field coupled oscillators.
 
The feedback mechanism is considered as one of the main themes of scientific understanding of the last century and has been widely used in control theory \cite{Astrom_book}. The positive feedback which favors the system's instability in the dynamical system has been used extensively in neural networks, genetic networks, etc\cite{Attila_Becskei}. On the other hand, negative feedback promotes stability and has been widely used to model biochemical systems\cite{Nguyen}.  It is one of the fundamental mechanisms in cellular networks and is shown to be present in many biochemical systems including bacterial adaptation\cite{Kollmann}, mammalian cell cycle\cite{Ferrel}, etc. But the negative feedback with time delay favours oscillatory dynamics in the system\cite{Borsch2016}, a scenario that we  explore in this work as well.

It is a well-established fact that time delays are an essential part of many natural systems. In recent years, more and more systems are being recognised to be influenced by or to be describable via a delayed coupling. Time delay due to the finite propagation speed of the external signal, which is described as propagation delay has been widely used to control the dynamics of coupled nonlinear systems\cite{PhysRevE.80.065204,PhysRevE.86.036210}. The internal self-feedback delay appears because sometimes the system needs a finite time to process the received signal and then act on it. It is demonstrated that such type of local self-feedback delay acts as negative feedback and plays a crucial role in reviving oscillations or amplitude death and oscillation death \cite{Chaos.27.061101}. In comparison to propagation delay, the effects of self-feedback delay in coupling are very less explored. There are only a few instances where systems with self-feedback delay have been investigated.

In the context of  network of coupled oscillators feedback has been widely used for control of network dynamics and synchronization\cite{Bidesh2017,Chandrasekar2010}. However, the effects of feedback in the aging transition have not been well explored. Only recently, Kundu et al. \cite{Srilena1} bring out an in-depth study on the effects of external positive feedback to increase the dynamical persistence of a network of oscillators. In this work, we study the ability of negative self-feedback with a time delay to enhance the dynamical persistence of a network that is experiencing an aging transition. We have shown that delayed self-feedback is an effective control mechanism to enhance the dynamical robustness in a mean-field coupled network. We have shown that self-feedback delay very effectively enhances the robustness for global as well as local mean-field coupled network. Even when the amount of the local self-feedback delays is minimal, it effectively enhances the dynamical robustness of the network. We present an in-depth study of the global network.  For a global network to show that our enhancement mechanism is independent of the model, we provide results for the Stuart-Landau limit cycle system as well as the chaotic R\"ossler system. We elucidate our results, both analytically and numerically. To show that our method is independent of coupling topology, we also present numerical results for mean-field coupled oscillators interacting via complex network topology.

\section{Global Mean-Field Coupled Network}
In this section, we present an extensive study of the effect of negative self-feedback with a time delay to elevate the dynamical robustness in a global mean-field coupled network. Here global coupling signifies that all the dynamical units have equal contribution to the mean-field, which acts equally on all the units. We consider N mean-field coupled Stuart-Landau oscillators with timed delayed self-feedback. Mathematically one can write the governing equation of motion as
\begin{eqnarray}
\dot{z_j}(t)  &=& (\rho_j+i\omega-|z_j(t)|^2)z_j(t)+k[\overline{z}-z_j(t-\tau)],
\label{e1}
\end{eqnarray}

\noindent for $j = 1,2, . . . ,N$. Where $z_j=x_j+iy_j$ is the complex amplitude of the $j$th oscillator, and $\overline{z}$ is the average mean-field. $\tau$ accounts for the delay in the local self-feedback term $z_j(t-\tau)$ , which basically acts as negative feedback. Here $\omega (=5)$ is the internal frequency of each oscillators, and $k$ for the coupling strength. $\rho_j$ is the bifurcation parameter of the $j$ th oscillator and gives rise to a supercritical Hopf bifurcation at $\rho_j=0$.  Each individual Stuart-Landau oscillator displays a stable sinusoidal type oscillation for $\rho_j=a (>0)$ and converges to stable equilibrium point $z_j = 0$ for $\rho_j=b (<0)$.

\begin{figure}[ht]
\centering
\includegraphics[width=0.49\textwidth]{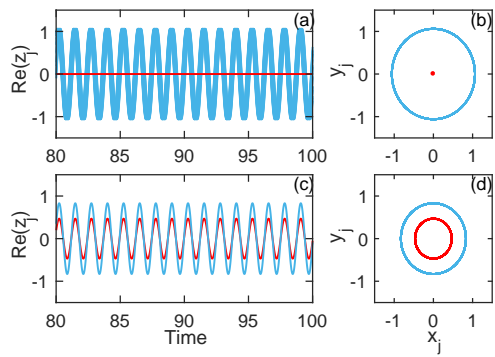}
\caption{Time series of real part of $z_j$ and phase portrait of  $N$ mean-field coupled Stuart-Landau oscillators consisting of active and inactive nodes are plotted  with (a,b) $k=0$ and (c,d) $k=2$ at the inactivation ratio $p=0.2$. The red color shows the inactive oscillators, and blue color indicates the active node of oscillators in the network.}
\label{f1}
\end{figure}


Here aging refers to the incidence when an active oscillator with $\rho_j = a > 0$ turns into inactive mode with $\rho_j = b < 0$ through some perturbations.  Without loss of generality, one can set the group of oscillators  ${j=1,2,.....N(1-p)}$  as active ones and the remaining oscillators  ${j=N(1-p)+1,.....N}$ as inactive. Here $N$ is the total number of oscillators in the network, and $p$ is the fraction of inactive oscillators. As the inactivation ratio $p$ reaches a critical value, the global oscillation of the network dies out.  Following Daido and Nakanishi\cite{daido2004}, we define an  order parameter $Z = \frac{|Z(p)|}{|Z(0)|}$ to quantify the the dynamics of the system, that defines the average magnitude of global oscillation in the network, where $|Z(p)| = N^{-1}\sum_{l=1}^N |z_l|$. We have taken the network size $N=500$ (results are also valid for larger network size) and consider the value of $a=1$ for active, and $b=-3$ for inactive oscillators. In the absence of any coupling ($k=0$), the dynamics of all the nodes divide into two groups of phase synchronized oscillation and steady-state.  For a certain value of $p=0.2$ the time series and phase portrait of $z_j$ is shown in the Fig.\ref{f1}(a-b). As we incorporate the coupling in the network, the inactive oscillators start oscillation under the influence of active oscillators. Both groups of oscillators show the synchronized motion with a different amplitude which is shown in Fig.\ref{f1}(c-d). In Figure. \ref{f2}(a) we have plotted $Z$ against the inactivation ratio $p$ for different values of local self-feedback delay $\tau$  for a fixed coupling strength $k=5$. One can observe that in the absence of local self-feedback delay $\tau$, the order parameter $Z$ vanishes at a lower value of $p=p_c$. This implies that the aging transition takes place much faster. As we increase the local self-feedback delay $\tau$, the aging transition can be observed at a higher value of $p=p_c$. So one can conclude that local delayed self-feedback plays an important role in enhancing the dynamical robustness. For a better understanding of the effect of local self delay feedback in the coupled oscillators, we have shown in Fig.\ref{f2}(b) the phase transition diagram in the plane $\tau-p$ for fixed $k=5$. In this figure, region OS and AT denote the oscillatory state and death state (where $Z=0$), respectively. It is evident from the figure that when $\tau$ is a minimal change in the $p_c$ value insignificant. But  as we increase the  value of $\tau$ on higher side, the critical value of $p_c$  increases and reach to the $p_c=1$ for $\tau=0.12$. It shows that the local self-feedback delay $\tau$ dominates the aging transition in the coupled oscillator and enhances the dynamical robustness of mean-field coupled oscillators.

\begin{figure}[t]
\centering
\includegraphics[width=0.49\textwidth]{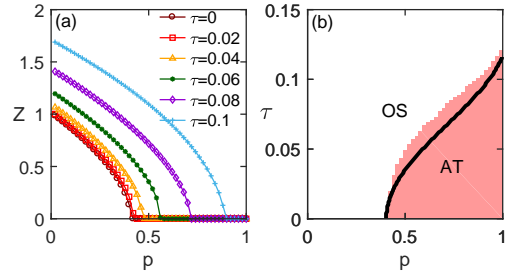}
\caption{(a) The order parameter $Z$ as a function of the inactivation ratio $p$ for various values of $\tau$ in a mean-field coupled network of $N=500$ Stuart-Landau oscillators for $a=1, b=-3, \omega=5$ and $k=5$. (b) The phase diagram in of coupled oscillator in ($p-\tau$) parameter plane at fixed $k=5$, where white and light pink region belongs to oscillatory state (OS) and aging transition (AT) region respectively. Solid black line is a fitting of the critical value of aging transition $p_c$ obtained from Eq.\ref{e6}.}
\label{f2}
\end{figure}

Next, we find the critical value $p_c$ analytically. The global oscillation collapses at $p_c$ during aging transition and the trivial fixed point $z_j = 0$ is stabilized. We assume that the coupled system build of two groups, where all nodes are identical in each group (basically, synchronized activity among the oscillators permits us to reformulate the system in such a way). By setting $z_j = A$ for the active group and $z_j =I$ for the inactive group of oscillators, the original equation \ref{e1} reduces to the following coupled systems \cite{daido2004},

\begin{eqnarray}
\dot{A(t)}&=&(a+i\omega+kq-|A(t)|^2)A(t)-kA(t-\tau)+kpI(t),\nonumber\\
\dot{I(t)}&=&(b+i\omega+kp-|I(t)|^2)I(t)-kI(t-\tau)+kqA(t),\nonumber\\
\label{e2}
\end{eqnarray}
\noindent where $q = 1-p$. Now, we cary out linear stability analysis to reduce Eq.\ref{e2} around the trivial fixed point $A = I = 0$. In order to find the critical value of the $p_c$. The origin $(A = I = 0)$ is stabilized if the real parts of the eigenvalues  become negative. Therefore, the critical inactivation ratio $p_c$ can be derived when two complex conjugate eigenvalues of above equations intersect the imaginary axis. A linear stability analysis around the origin now gives a characteristic equation of the form,
\begin{align}
(a+i\omega+qk-ke^{-\lambda\tau}-\lambda)\nonumber\\
(b+i\omega+pk-ke^{-\lambda\tau}-\lambda)
-pqk^2=0, 
\label{e3}
\end{align}
\noindent Here, $\lambda=\lambda_R+i\lambda_I$. By taking the real part of the eigenvalue equal to zero ($\lambda_R=0$) and separate the real and imaginary part of Eqs.~\ref{e3}, we obtain the following equations,
\begin{align}
[a+qk-k cos(\lambda_I\tau)][b+pk-k cos(\lambda_I\tau)]\nonumber\\
=(\omega-\lambda_I+k sin(\lambda_I\tau))^2+pqk^2,
\label{e4}
\end{align}
\begin{equation}
[a+b+k(p+q)-2k cos(\lambda_I\tau)][\omega-\lambda_I+k sin(\lambda_I\tau)] = 0,
\label{e5}
\end{equation}
\noindent where $p+q = 1$. Solving the above equations, we get critical value of inactivation ratio $p_c$ as, 
\begin{equation}
p_c = \frac{-ab+k(a+b)\beta+k^2\beta-kb-k^2\beta^2}{k(a-b)},
\label{e6}
\end{equation}
\noindent where $\beta = cos(\alpha\tau)$ and $\alpha = \omega+k\sqrt{1-\left(\frac{a+b+k}{2k}\right)^2}$.

The occurrence of the aging transition is featured by the existence of a critical parameter $p_c$. The critical value of the $p_c$ (black solid line) also has a good agreement with the numerical result (shaded region) for the aging transition shown in Fig.~\ref{f2}(b). 


\begin{figure}[t]
\includegraphics[width=0.49\textwidth]{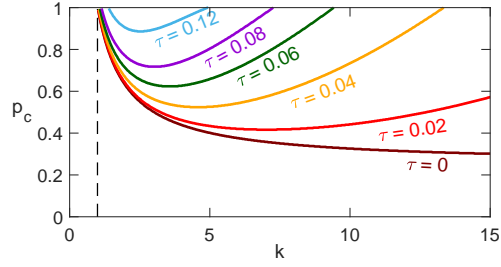}
\caption{The critical value of aging transition $p_c$ as a function of coupling strength $k$ for different values of local self-delay $\tau$ in the mean-field coupled oscillators, where $p_c$ monotonically decreases for increasing of $k$.}
\label{f3}
\end{figure}
When the  local self-feedback delay $\tau=0$, the aging transition takes place  for all $k>1$ and $p_c$ decreases as coupling strength $k$ increases. To explore the impact of $k$ on the parameter $p_c$, we plot the $p_c$ as a function of coupling strength $k$ for different $\tau$ values in Fig.~\ref{f3}. Surprisingly, we observe that for nonzero  $\tau$, the aging transition exists only for a finite interval of coupling strength $k$. Within this range, $p_c$ gradually decreases till it reaches its minimum value and then again monotonically increases to unity. The above observation implies that strong coupling favors dynamical resilience of the network against aging when feedback delay $\tau$ is large enough.


To demonstrate that our enhancement method is independent of the dynamical model, we choose a chaotic dynamical system coupled through mean-field with local self-delay feedback. The governing equation  of $N$ mean-field  coupled R\"ossler oscillators  with negative self-feedback  delay can be written as:

\begin{eqnarray}
\dot{x}_j (t)&=& -y_j(t)-z_j(t)+k[\overline{x}-x_j(t-\tau)],\nonumber\\
\dot{y}_j (t)&=& x_j(t)+r_jy_j(t)+k[\overline{y}-y_j(t-\tau)],\nonumber\\
\dot{z}_j(t) &=& r_j+z_j(t)(x_j(t)-e)+k[\overline{z}-z_j(t-\tau)],
\label{e7}
\end{eqnarray}
\noindent where $j=1,2,...N$. $r_j, e (=5.7)$ are the intrinsic parameters of R\"ossler oscillator. $\overline{x}, \overline{y}$, and $\overline{z}$ are the meanfield term of $x, y, z$ variables. We set the $r_j=-0.2$ for $j=1,...,pN$ for the inactive oscillators falls into a fixed point and $r_j=0.2$ for $j=pN+1,...,N$ for active oscillators which shows the the chaotic oscillations. To study the aging transition in chaotic oscillators, we measure the amplitude of oscillation $R=\frac{M(p)}{M(0)}$, where $M$ is define as \cite{daido2004},
\begin{equation}
	M = \sqrt{\langle(X_c-\langle X_c\rangle)^2\rangle},
\end{equation}
\noindent $X_c=N^{-1}\sum_{j=1}^N(x_j,y_j,z_j)$ is the centroid, and $\langle .\rangle$ means it calculated for long time average.


\begin{figure}[t]
\centering
\includegraphics[width=0.49\textwidth]{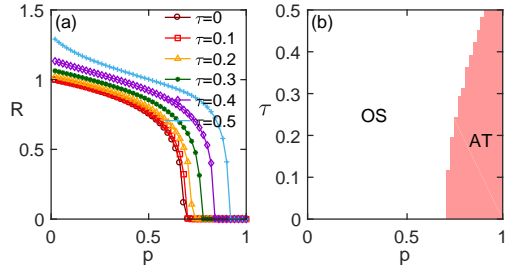}
\caption{(a) The order parameter $R$ is plotted as a function of the inactivation ratio $p$ at the various values of $\tau$ in mean-field coupled network of $N=500$ R\"ossler oscillators for $e=5.7$ and $k=0.2$. (b) The phase diagram in of coupled oscillators in ($p-\tau$) parameter plane at fixed $k=0.2$, where white and light pink region belongs to oscillatory state (OS) and aging transition (AT) region.}
\label{f4}
\end{figure}

\begin{figure}[t]
\centering
\includegraphics[width=0.49\textwidth]{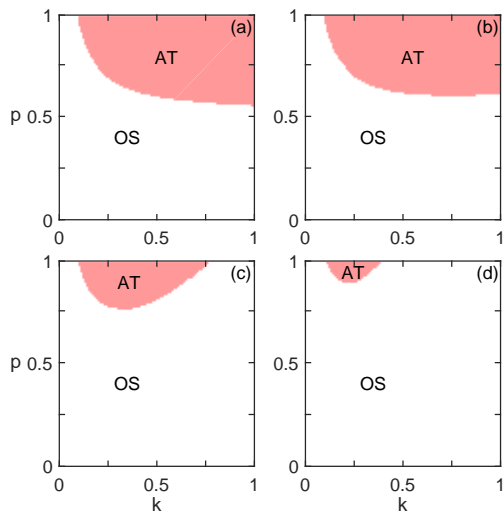}
\caption{The phase diagram in ($p, k$) parameter plane for the network of mean-field coupled R\"ossler oscillators. The aging transition (AT) and oscillatory state (OS) regions are characterized by the light pink and white color for different value of local self-delay (a) $\tau=0$, (b) $\tau=0.2$, (c) $\tau=0.4$, and (d) $\tau=0.6$.}
\label{f5}
\end{figure}

We have plotted the order parameter $R$ in Figure~\ref{f4}(a) against  the  inactivation ratio $p$ for the fixed value $k=0.2$ at different values of $\tau$. In the absence of local self-delay $\tau=0$, the brown line of $R$ demarcated the critical aging transition point at $p_c=0.72$, and when $\tau$ is increased, the transition threshold point $p_c$ increase to the higher value.  It shows that local delayed self-feedback effectively enhances the robustness.  In Fig.~\ref{f4}(b), we have shown the phase diagram in $\tau-p$ parameter plane for a fixed coupling strength $k=0.2$, where two regions are observed, oscillatory state (OS) and aging transition (AT). The critical value $p_c$ of aging transition from oscillatory to steady-state (where $R=0$) increases with $\tau$ gradually.  For R\"ossler model also we investigate the effects of coupling strength on robustness. We have shown the phase diagram in ($p-k$) parameter space for different value of $\tau$ in Figure \ref{f5}. The light pink region in the parameter plane shows the aging transition where the order parameter $R$ falls to zero. It clearly shows that the AT region decreases in size for the higher values of $\tau$, which is quite similar to the results we have obtained for Stuart-Landau oscillators. It confirms that self-feedback delay $\tau$ is also very useful in inflating the dynamical robustness of a coupled chaotic oscillator network.

\section{ Complex Network Topology}
Next, we demonstrate the effectiveness of our enhancement technique when the oscillators are interacting through a complex network topology\cite{Osipov2007, Amit@PLA2019}. We Consider $N$ mean-field coupled  Stuart-Landau oscillators  interacting via complex network topology.  The mathematical model of the coupled system is given by
\begin{eqnarray}
\dot{z_j (t)}&=& (\rho_j+i\omega-|z_j(t)|^2)z_j(t)+k \left(\frac{\sum_{l=1}^N A_{jl}z_l(t)}{ r_j}-z_j(t-\tau)\right),\nonumber\\
\label{e10}
\end{eqnarray}
\noindent for $j = 1,2, . . . ,N$ and $N=500$.  Here $A_{jl}$ is the adjacency matrix of the connection in the complex network, i.e., $A_{jl}= 1$ if $j$-th and $l$-th nodes are connected and zero otherwise. Here, we assume that all coupled oscillators are interacting with each other through specific network topology and all the connection between them have the same coupling strength. $r_j$ is the degree of node j, and $k$ is the coupling strength of interaction.

Depending on the probability distribution of the degrees of nodes, we can categorize complex networks into two broad groups, namely homogeneous networks and heterogeneous networks \cite{Barabasi}. Degree distribution of homogeneous networks such as random graphs follows a binomial or Poisson distribution. In contrast, heterogeneous networks such as scale-free networks obey a heavy-tailed degree distribution that can be approximated by a power law ($P(k)\approx k^{-\eta}$, where $P(k)$ is the probability of having a node of degree $m$ and $\eta$ is the power-law exponent).

In Fig.~\ref{f6}(a), we have plotted the order parameter $Z$ as a function of inactivation ratio $p$ for different value of self-feedback delay term $\tau$ for a random network of $N=500$ and the probability of connecting two distinct pair of nodes $\gamma=0.02$. It demonstrates the fact the $p_c$ value increases as we increase $\tau$.  Even for a small amount of delay $\tau$, dynamical robustness gets enhanced to a significant amount. We have similar results for the scale-free network. In Fig.~\ref{f6}(b), we have shown how $p_c$ value increases as we increase $\tau$. These outcomes show that local self-feedback with time delay is very efficient in enhancing the dynamical robustness when the oscillators are interacting through a complex network topology. It also establishes the fact that our enhancement technique is independent of coupling topology.

\begin{figure}[t]
\centering
\includegraphics[width=0.49\textwidth]{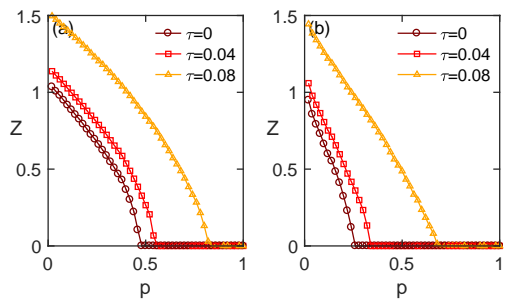}
\caption{The order parameter $Z$ as a function of the inactivation ratio $p$ for various values of $\tau$ in complex network of $N=500$  Stuart-Landau oscillators for $a=1, b=-3, \omega=5$ and $k=5$ for (a) random network for probability of connecting two distinct pair of nodes $\gamma=0.02$, and (b) scale-free network  of average degree $\langle m\rangle$=8, and the the power law exponent $\eta=3.0$ .}
\label{f6}
\end{figure}

\section{Conclusion}

In this paper, we have demonstrated that the introduction of a negative self-feedback with a delay into the mean-field coupled oscillators can effectively increase the dynamical robustness, which is exhibited by enhancing the endurance of the oscillatory dynamics of the network against  the aging of the individual nodes. We have demonstrated that our enhancement technique is applicable for both global as well as local mean-field coupling.  We have shown that the critical vale $p_c$ at which the aging transition takes place is positively correlated with the delay term $\tau$. It means the network becomes more resilient to the effect of the local deterioration of oscillatory nodes with increases of $\tau$. 

We have done a comprehensive, detailed study for the global network. We have seen that the enhancing effect of negative self-feedback with delay is more prominent for a strong coupling term $k$. 
To our surprise, we have found that for a non zero $\tau$ the aging transition exists for a finite length of the coupling strength k. Within this range, $p_c$ first diminishes from maximum to a minimum value and then again increases to the maximum. The global oscillations can be restored by local self-feedback with delays in coupled networks of purely non-oscillatory units. The introduction of delayed self-feedback in the coupling provides a straightforward but highly valuable technique for recovering dynamical activities in the network, whose oscillatory behavior has been weakened due to the aging of some elements. We have delineated the results using numerical simulations as well as analytical findings. To state that our scheme is independent of local subsystems, we have successfully employed our control mechanism to a network of coupled chaotic R\"ossler oscillators.

We have also studied the role of negative self-feedback with a time delay to enhance dynamical robustness when the oscillators interact through a complex network topology.  The results demonstrate the fact that our enhancement technique is independent of coupling topology.

Our study might have applicability in enhancing the resilience of several natural systems that can experience aging in its local dynamical units.  Lastly, our proposed framework  widens the understanding of the roles of negative self-feedback delay in regulating oscillatory dynamics to design network  oscillators which can resist the aging transition.} 

\section{DATA AVAILABILITY}
The data that support the findings of this study are available from the corresponding author upon reasonable request.
\section{Bibliography}
\bibliographystyle{spphys}       
\bibliography{ref_robust}   

\end{document}